\documentclass[%
 aip,
rsi,%
 amsmath,amssymb,
 reprint,%
]{revtex4-1}

\usepackage{graphicx}
\usepackage{dcolumn}
\usepackage{bm}

\usepackage{epstopdf}
\usepackage[mathlines]{lineno}

\usepackage[
pdfauthor={},
pdftitle={},
bookmarks=false,
bookmarksnumbered,
pdftex,
pdfstartview=FitH 
]{hyperref}
\hypersetup{colorlinks=true, 
linkcolor=blue,
urlcolor=blue,
citecolor=blue}
\usepackage[all]{hypcap} 

\begin{document}

\title{Reconfigurable edge-state engineering in graphene using LaAlO$_3$/SrTiO$_3$ nanostructures}

\author{Jianan Li}
\altaffiliation{These authors contributed equally to this work.}
\affiliation{Department of Physics and Astronomy, University of Pittsburgh, Pittsburgh, PA 15260, USA}

\author{Qing Guo}
\altaffiliation{These authors contributed equally to this work.}
\affiliation{Department of Physics and Astronomy, University of Pittsburgh, Pittsburgh, PA 15260, USA}

\author{Lu Chen}
\affiliation{Department of Physics and Astronomy, University of Pittsburgh, Pittsburgh, PA 15260, USA}

\author{Shan Hao}
\affiliation{Department of Physics and Astronomy, University of Pittsburgh, Pittsburgh, PA 15260, USA}

\author{Yang Hu}
\affiliation{Department of Physics and Astronomy, University of Pittsburgh, Pittsburgh, PA 15260, USA}

\author{Jen-Feng Hsu}
\affiliation{Department of Physics and Astronomy, University of Pittsburgh, Pittsburgh, PA 15260, USA}

\author{Hyungwoo Lee}
\affiliation{Department of Materials Science and Engineering, University of Wisconsin-Madison, Madison, WI 53706, USA}

\author{Jung-Woo Lee}
\affiliation{Department of Materials Science and Engineering, University of Wisconsin-Madison, Madison, WI 53706, USA}

\author{Chang-Beom Eom}
\affiliation{Department of Materials Science and Engineering, University of Wisconsin-Madison, Madison, WI 53706, USA}

\author{Brian D'Urso}
\affiliation{Department of Physics, Montana State University, Bozeman, MT 59717, USA}

\author{Patrick Irvin}
\affiliation{Department of Physics and Astronomy, University of Pittsburgh, Pittsburgh, PA 15260, USA}

\author{Jeremy Levy}
\email{jlevy@pitt.edu}
\affiliation{Department of Physics and Astronomy, University of Pittsburgh, Pittsburgh, PA 15260, USA}

\date{\today}

\begin{abstract}
The properties of graphene depend sensitively on doping with respect to the charge-neutrality point (CNP). Tuning the CNP usually requires electrical gating or chemical doping. Here, we describe a technique to reversibly control the CNP in graphene with nanoscale precision, utilizing LaAlO$_3$/SrTiO$_3$ (LAO/STO) heterostructures and conductive atomic force microscope (c-AFM) lithography. The local electron density and resulting conductivity of the LAO/STO interface can be patterned with a conductive AFM tip \cite{cen2008nanoscale} , and placed within two nanometers of an active graphene device \cite{huang2015electric} . The proximal LAO/STO nanostructures shift the position of graphene CNP by $\sim 10^{12}$ cm$^{-2}$, and are also gateable. Here we use this effect to create reconfigurable edge states in graphene, which are probed using the quantum Hall effect. Quantized resistance plateaus at $h/e^2$ and $h/3e^2$ are observed in a split Hall device, demonstrating edge transport along the c-AFM written edge that depends on the polarity of both the magnetic field and direction of currents. This technique can be readily extended to other device geometries. 
\end{abstract}

\maketitle

Graphene has proved to be a powerful and versatile platform for studying condensed matter phenomena due to the unique honeycomb crystal structure and Dirac fermion behavior of electrons. The unique Dirac cone band structure makes it possible to tune the carrier density continuously between electrons and holes. This duality of carriers in graphene results in many exotic properties of graphene, such as Klein tunneling\cite{allain2011klein, katsnelson2006chiral, young2009quantum, shytov2008klein}, edge state mixing\cite{williams2007quantum, abanin2007quantized, lohmann2009four, amet2014selective}, and recently the ``wedding cake'' structure of quantum Hall states\cite{gutierrez2018interaction}. Central to many of these experimental findings is the ability to control the charge neutrality point (CNP) by electrical gating. 

Another well-studied two-dimensional electronic system is the LaAlO$_3$/SrTiO$_3$ (LAO/STO) heterostructure, which supports a high mobility 2D electron layer\cite{ohtomo2004high} with a wide range of additional properties including magnetism\cite{brinkman2007magnetic}, tunable spin-orbit coupling\cite{caviglia2010tunable, fischer2013spin, shalom2010tuning}, superconductivity\cite{reyren2007superconducting}, and BEC-like superconductivity\cite{cheng2015electron}. The 2DEG on the interface is globally tunable with a back-gate voltage and locally tunable from the top LAO surface using conductive atomic force microscope (c-AFM) tip, when the LAO thickness is close to a critical thickness $\sim$3 unit cells\cite{cen2008nanoscale, thiel2006tunable, cen2009oxide}. Using c-AFM lithography, a wide range of devices on LAO/STO interface can be fabricated, such as a single electron transistor\cite{cheng2011sketched}, broadband terahertz source and detector\cite{jnawali2015photoconductive, chen2019over},  one-dimensional interference device\cite{annadi2018quantized, pai2018one} and electron waveguide\cite{pai2018physics}. This technique can also be applied to other complex oxide heterostructures as well\cite{chen2018extreme}.

There have been efforts to locally control the CNP of graphene on silicon or hBN substrates using AFM\cite{schmidt2013mixing} or STM\cite{velasco2016nanoscale}. However, those doping techniques are either non-reversible, or can only be performed in ulrtra-high vacuum and low temperature, which limits the applications. In this work, we demonstrate how local control over the metal-insulator transition in LAO/STO can be used to reversibly pattern interacting edge channels in a proximal graphene layer under ambient conditions. The graphene used in this work is grown from chemical vapor deposition (CVD) on oxygen-free electronic grade copper flattened with a diamond turning machine\cite{dhingra2014chemical}. Then graphene is coated with perfluoropolymer Hyflon AD60 and then transferred onto the LAO/STO surface with wet-transfer technique\cite{li2016method}. Graphene is patterned into Hall bars by standard photolithography. Hyflon is removed from Graphene with FC-40 after patterning. Particles and contaminants on graphene from wet transfer and photolithography are brushed away using a contact-mode AFM scan sequence. After cleaning, the 4 $\mathrm{\AA}$ atomic steps of the LAO surface underneath graphene are clearly resolvable\cite{li2016method}. The quality of the graphene is similar to other samples prepared in similar methods, with the mobility $\mu > 10,000$ cm$^{2}$V$^{-1}$s$^{-1}$ at 2 K\cite{li2016method}.

Figure 1(a,b) illustrate the c-AFM writing setup. Graphene is scanned with a conductive doped-silicon tip in contact mode with a contact force of 15 $\sim$ 20 nN and scanning speed between 1 $\mu$m/s and 10 $\mu$m/s . The bias voltage applied on the tip is set to $+17$ V (for creating a conductive LAO/STO interface) or $-5$ V (for restoring an insulating LAO/STO interface while avoiding damage to graphene\cite{alaboson2011conductive}). After each raster scan of the graphene area, the CNP of the graphene in the written region is shifted. The mechanism for shifting the CNP is believed to be essentially the same as for tuning the LAO/STO interface without graphene\cite{huang2015electric, li2016method, bi2010water}. Under ambient conditions, when a positive voltage is applied to the tip while graphene is grounded, water molecules adsorbed on the graphene surface are dissociated into protons, transferred through the graphene and mediate the metal-insulator transition in the LAO/STO while contribute to a shift in the chemical potential in the graphene layer\cite{huang2015electric, bi2010water, brown2016giant, hu2014proton}. The CNP can be further shifted by dynamically changing the electron density in the LAO/STO layer.

STO has high dielectric permittivity at low temperature ($\epsilon_r \sim 10,000$)\cite{weaver1959dielectric}, which enables the graphene to be easily tuned with a back-gate voltage $V_\mathrm{bg}$ applied to the bottom of the LAO/STO substrate (Figure 1(b)). However, this gating method is subject to significant hysteresis\cite{couto2011transport, jnawali2017room} (see Fig. S1(a), inset), and hence the back-gate voltage is not a reliable indicator of doping level with respect to the CNP.  In addition, the c-AFM lithography itself will dope the graphene, even when the back-gate voltage is held fixed. For these reasons, we rely on the four-terminal resistance of the graphene to monitor the carrier density change \textit{in situ} during the c-AFM writing process, which takes place under the condition $V_\mathrm{bg} = 0$ V (more details are discussed in the supplementary materials). Once the c-AFM writing is finished, the sample is immediately stored in vacuum and cooled to cryogenic temperatures, where the writing is known to persist indefinitely\cite{bi2010water, huang2015electric}. To directly illustrate the effect of c-AFM writing, we scan half of the graphene device with $V_\mathrm{tip} = +17$ V, as shown in Figure 1(c). The graphene resistance is then measured as a function of back-gate voltage at $T = 2$ K. Figure 1(d) shows a control measurement where the resistance is measured before c-AFM scanning. The peak at $V_\mathrm{bg} = 5$ V clearly indicates the CNP. Figure 1(e) is measured after c-AFM writing shown in Figure 1(c), and two peaks can be observed. The additional peak on the left-hand-side is attributed to the c-AFM writing.

The graphene doping from the positively biased c-AFM tip is reversible. After the c-AFM writing and the change in four-terminal resistance being observed, a scan with $V_\mathrm{tip}= -5$ V voltage on the c-AFM tip will partially remove the previous writing effect. Scans with negative $V_\mathrm{tip}$ need to be carefully conducted and the c-AFM tip should be connected in series with a 1 G$\Omega$ resistor, due to the fact that graphene can be oxidized as anode\cite{alaboson2011conductive, byun2011nanoscale}. Also, graphene has to be detached from measurement leads or groundings so that there is no significant current flowing through graphene\cite{alaboson2011conductive}. 

The carrier density in the LAO/STO-doped graphene device is quantified using the Hall effect. As shown in the inset of Figure 2(a), a graphene/LAO/STO device is prepared with one graphene Hall cross (Hall B) scanned under contact mode with the c-AFM tip biased at $+17$ V for 15 times. A second Hall cross device (Hall A) is measured as a control, where no c-AFM lithography is performed. An electrical gate connected to the back of the 1 mm thick STO substrate is used to adjust the overall carrier density of the graphene device. Magnetotransport experiments are performed at $T = 2$ K, in an out-of-plane magnetic field ($B = 1$ T), in order to determine the carrier densities of the two regions. A shift of $\Delta n = 7 \times 10^{11}$ cm$^{-2}$ is observed, with the patterned area being more $n$-type. Because Hall Device B is locally gated positively, the CNP is shifted to a lower $V_\mathrm{bg}$ value (green curve). The carrier densities on both regions can be tuned by the back-gate up to $1 \times 10^{13}$ cm$^{-2}$ at $V_\mathrm{bg} = -10$ V, in part due to the large dielectric constant of STO ($\epsilon_r \sim 10,000$) at 2 K\cite{weaver1959dielectric, couto2011transport}. The right ends of the curves are less linear and tends to be saturated, due to the shielding effect of the 2DEG on LAO/STO interface induced by a high positive back-gate voltage. For $V_\mathrm{bg} < 5$ V the interface of LAO/STO outside the previously written area is insulating, so the back-gate voltage will not be shielded. 

At sufficiently large magnetic fields, graphene would exhibit quantized Hall resistance of $R_\mathrm{h} = h/[(4n+2)e^2] \ (n = 0, 1, 2 \ldots)$ and vanishing longitudinal resistance, as a result of the non-trivial Berry phase and four-fold degeneracy from electron spin and valley pseudo-spin\cite{novoselov2004electric, geim2010rise, zhang2005experimental}. When the two adjacent regions have different Landau level filling factors, for example a $p$-$n$ junction in the quantum Hall regime\cite{williams2007quantum, amet2014selective}, the mixing and equilibration of edge states will produce a non-zero longitudinal resistance, which follows Landauer-Buttiker formalism\cite{buttiker1986four, buttiker1988absence}. In our sample, the $\Delta n = 7 \times 10^{11}$ cm$^{-2}$ carrier density difference on the two sides is enough to keep them at adjacent Landau level filling factors. Consequently, these two regions have different edge-channel occupancies. As shown in Figure 2(b), when both regions have the same polarity, the channels present in both regions would travel across both regions, while those ones from higher filling factors would only circulate in one region. The longitudinal resistance $R_\mathrm{xx1}$ and $R_\mathrm{xx2}$ measured from the top and bottom of the sample can be described using the Landauer-Buttiker formalism\cite{williams2007quantum, abanin2007quantized, amet2014selective, buttiker1986four, buttiker1988absence, ki2010dependence, klimov2015edge, woszczyna2011graphene, ozyilmaz2007electronic, schmidt2013mixing} (details of derivations can be found in the supplementary materials):
\[
R_\mathrm{xx1} = \frac{h}{e^2} \left(\frac{1}{|\nu_2|} - \frac{1}{|\nu_1|} \right), \ \ R_\mathrm{xx2} = 0,
\]
where $\nu_1$ and $\nu_2$ are the filling factors of the two regions, equal $\pm2, \pm6, \ldots$. In the case of opposite polarity on two sides the device becomes a $p$-$n$ junction and the current flows in opposite directions on each side. The longitudinal resistance $R_\mathrm{xx1}$ and $R_\mathrm{xx2}$ become:
\[
R_\mathrm{xx1} = 0, \ \ R_\mathrm{xx2} = \frac{h}{e^2} \left(\frac{1}{|\nu_1|} + \frac{1}{|\nu_2|}\right).
\]

Figure 3(a) and 3(b) shows $R_\mathrm{xx1}$ and $R_\mathrm{xx2}$ in $+7$ T and $-7$ T magnetic fields. When the back gate is swept from $-10$ V to $+10$ V, the carrier type in the two regions would transit from unipolar ($\nu_1 = -6$, $\nu_2 = -2$) to bipolar ($\nu_1 = -2$, $\nu_2 = +2$) and then unipolar ($\nu_1 = +2$, $\nu_2 = +6$) again. As shown in Figure 3(a), when the back-gate voltage $V_\mathrm{bg}$ is between $-2$ V and $+6$ V, the resistance $R_\mathrm{xx1}$ transitions from $h/3e^2$ to 0 and then to $h/3e^2$, while $R_\mathrm{xx2}$ transitions from 0 to $h/e^2$ and then to 0, as predicted by the Landauer-Buttiker formalism. When the magnetic field is reversed, the quantization of $R_\mathrm{xx1}$ and $R_\mathrm{xx2}$ are switched, because of the reversing of current directions. Figure 3(c) and (d) shows the swapping of quantization between $R_\mathrm{xx1}$ and $R_\mathrm{xx2}$ when the magnetic field is reversed. These results are consistent with the graphene edge-state mixing reported in literature\cite{schmidt2013mixing, lohmann2009four}. The values of resistance plateaus are quite close to theoretical values when the magnetic field is higher than 2 T, suggesting well-defined edge states equilibrium. The quantization features experience negligible change over the course of the measurement ($>$ 10 hours), indicating that the graphene doping is stable in vacuum, similar to the c-AFM writing on bare LAO/STO\cite{bi2010water}. 

In summary, we developed a reversible, spatially controllable graphene doping technique by c-AFM tips on LAO/STO substrates. Graphene edge state mixing in quantum Hall regime can be observed from with the c-AFM writing. In the future, this technique can be used locally dope high-mobility graphene with feature sizes as small as 20 nm \cite{huang2015electric}, to create a new family of reconfigurable graphene metamaterials.

This research was supported by the Office of Naval Research (N00014-16-1-3152), National Science Foundation and US-Israel Binational Science Foundation (DMR-1609519), and Vannevar Bush Faculty Fellowship program sponsored by the Basic Research Office of the ASD (R\&E) and funded by the ONR (N00014-15-1-2847). The work at University of Wisconsin-Madison was supported by the National Science Foundation (DMR-1629270) and the Air Force Office of Scientific Research (FA9550-15-1-0334).
\section*{Supplementary material}

See supplementary material for the details of hysteresis from back-gate voltage sweep, resistance and carrier density of graphene as functions of back-gate voltage, graphene resistance change during the c-AFM writing process, and the derivations of longitudinal resistances for edge-state mixing.

\begin{acknowledgments}
This research was supported by the Office of Naval Research (N00014-16-1-3152), National Science Foundation and US-Israel Binational Science Foundation (DMR-1609519), and Vannevar Bush Faculty Fellowship program sponsored by the Basic Research Office of the ASD(R\&E) and funded by the ONR (N00014-15-1-2847). The work at University of Wisconsin-Madison was supported by the National Science Foundation (DMR-1629270) and the Air Force Office of Scientific Research (FA9550-15-1-0334).
\end{acknowledgments}

\bibliography{bibliography}

\newpage

\begin{figure*}[tp]
\includegraphics[width=0.9\textwidth]{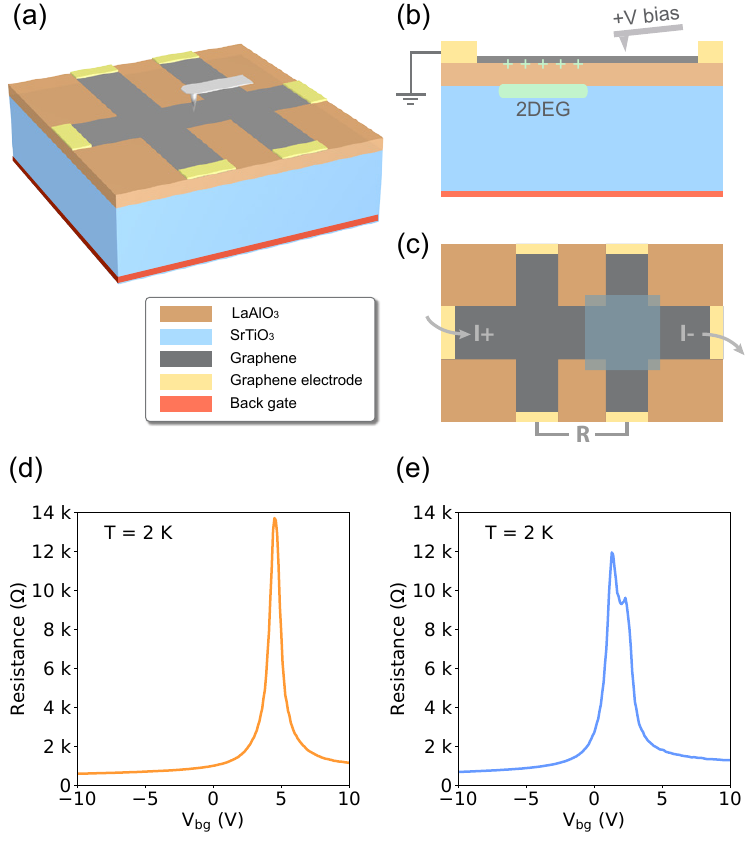}
\label{FIG1}
\caption{(a) The graphene (gray area) is transferred onto LAO/STO (orange and blue) surface and patterned into Hall bars. (b) In the ambient condition, a positively biased tip can dissociate protons from water molecules. Protons can permeate through graphene and reach LAO surface, shift the CNP of graphene locally and induce 2DEG on the interface of LAO/STO. (c) A square region covering half of the graphene device is scanned by a positively biased c-AFM tip, with $V_\mathrm{tip} = +17$ V. (d) The four-terminal resistance of the graphene device as a function of back-gate voltage, before c-AFM scanning with tip voltage. The measurement is performed at $T = 2$ K. The CNP is clearly seen at $V_\mathrm{Vb} = 5$ V. (e) Graphene resistance after c-AFM writing on part of the device. A clear splitting of CNP can be observed.}
\end{figure*}

\newpage

\begin{figure*}[tp]
\includegraphics[width=0.8\textwidth]{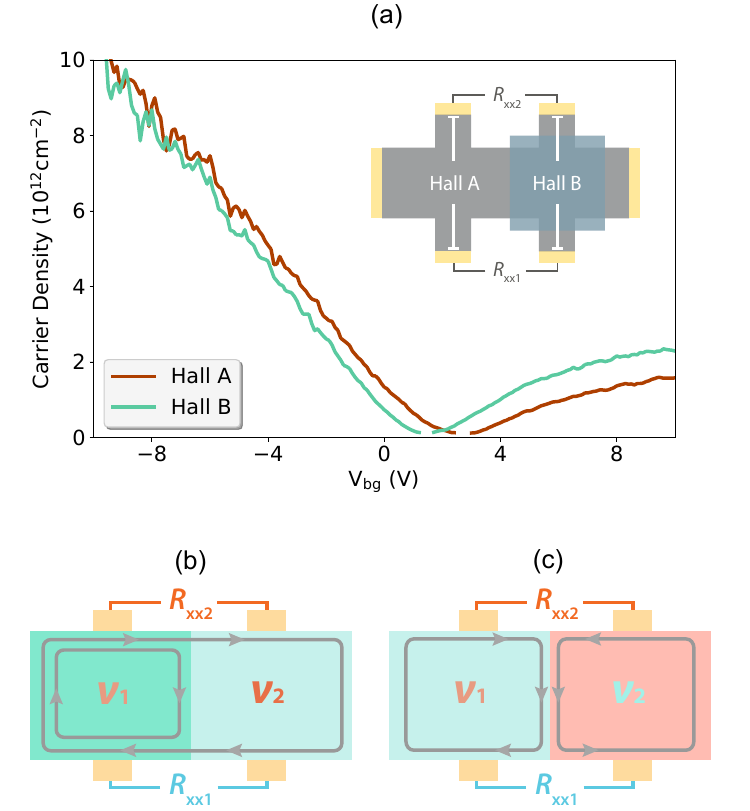}
\label{FIG2}
\caption{(a) The carrier density of graphene is measured at 2 K after doped with c-AFM scanning. Inset: Hall cross on the right-hand side (Hall B) is scanned several times with $+17$ V on the AFM tip, while the Hall cross on the left-hand side (Hall A) is measured as control. Carrier densities measured from Hall effect at 2 K shows that the CNP of Hall B is shifted to the left-hand side, as a result of doping with c-AFM. (b) When the two regions have the same polarity and $|\nu_1| \neq |\nu_2|$, the Landau level filling factors are different. Region on the left-hand side can support more edge channels. The edge states from higher filling factors are reflected on the boundary in the middle, and recombined with the state coming from the right on the bottom. (c) When the back-gate voltage is set at the value where the carriers are $n$- and $p$-type on each side, the directions of current are the opposite of each other. The mixing on the boundary will cause non-trivial quantization of longitudinal conductance.}
\end{figure*}

\newpage

\begin{figure*}[tp]
\includegraphics[width=0.8\textwidth]{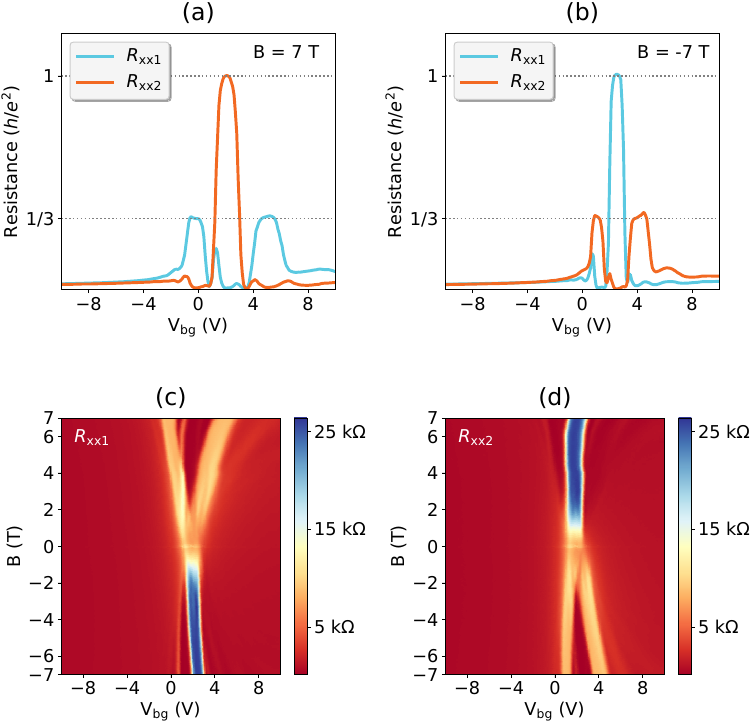}
\label{FIG3}
\caption{(a) and (b) are the resistance of $R_\mathrm{xx1}$ and $R_\mathrm{xx2}$ (as shown in Figure 2(b) and 2(c)) in the magnetic field of $+7$ T and $-7$ T. In (a)  $R_\mathrm{xx1}$ shows a plateau at $h/3e^2$ when the two regions are unipolar with filling factors $|\nu| = $ 2 and 6 respectively.  $R_\mathrm{xx2}$ shows a plateau at $h/e^2$ when the two regions are bipolar with filling factor $\nu = +2$ and $-2$. The plateaus of the two curves are anti-symmetric with respect to the magnetic field. When the direction of field is reversed, the resistance values and features are swapped between $R_\mathrm{xx1}$ and $R_\mathrm{xx2}$. (c) and (d) shows that the edge state mixing is well developed at $|B| > 2$ T.}
\end{figure*}

\end{document}


\title{Supplementary Materials for Reconfigurable edge-state engineering in graphene using LaAlO$_3$/SrTiO$_3$ nanostructures}

\author{Jianan Li}
\altaffiliation{These authors contributed equally to this work.}
\affiliation{Department of Physics and Astronomy, University of Pittsburgh, Pittsburgh, PA 15260, USA}

\author{Qing Guo}
\altaffiliation{These authors contributed equally to this work.}
\affiliation{Department of Physics and Astronomy, University of Pittsburgh, Pittsburgh, PA 15260, USA}

\author{Lu Chen}
\affiliation{Department of Physics and Astronomy, University of Pittsburgh, Pittsburgh, PA 15260, USA}

\author{Shan Hao}
\affiliation{Department of Physics and Astronomy, University of Pittsburgh, Pittsburgh, PA 15260, USA}

\author{Yang Hu}
\affiliation{Department of Physics and Astronomy, University of Pittsburgh, Pittsburgh, PA 15260, USA}

\author{Jen-Feng Hsu}
\affiliation{Department of Physics and Astronomy, University of Pittsburgh, Pittsburgh, PA 15260, USA}

\author{Hyungwoo Lee}
\affiliation{Department of Materials Science and Engineering, University of Wisconsin-Madison, Madison, WI 53706, USA}

\author{Jung-Woo Lee}
\affiliation{Department of Materials Science and Engineering, University of Wisconsin-Madison, Madison, WI 53706, USA}

\author{Chang-Beom Eom}
\affiliation{Department of Materials Science and Engineering, University of Wisconsin-Madison, Madison, WI 53706, USA}

\author{Brian D'Urso}
\affiliation{Department of Physics, Montana State University, Bozeman, MT 59717, USA}

\author{Patrick Irvin}
\affiliation{Department of Physics and Astronomy, University of Pittsburgh, Pittsburgh, PA 15260, USA}

\author{Jeremy Levy}
\email{jlevy@pitt.edu}
\affiliation{Department of Physics and Astronomy, University of Pittsburgh, Pittsburgh, PA 15260, USA}

\date{\today}

\maketitle

\begin{figure*}[tp]
\includegraphics[width=0.8\textwidth]{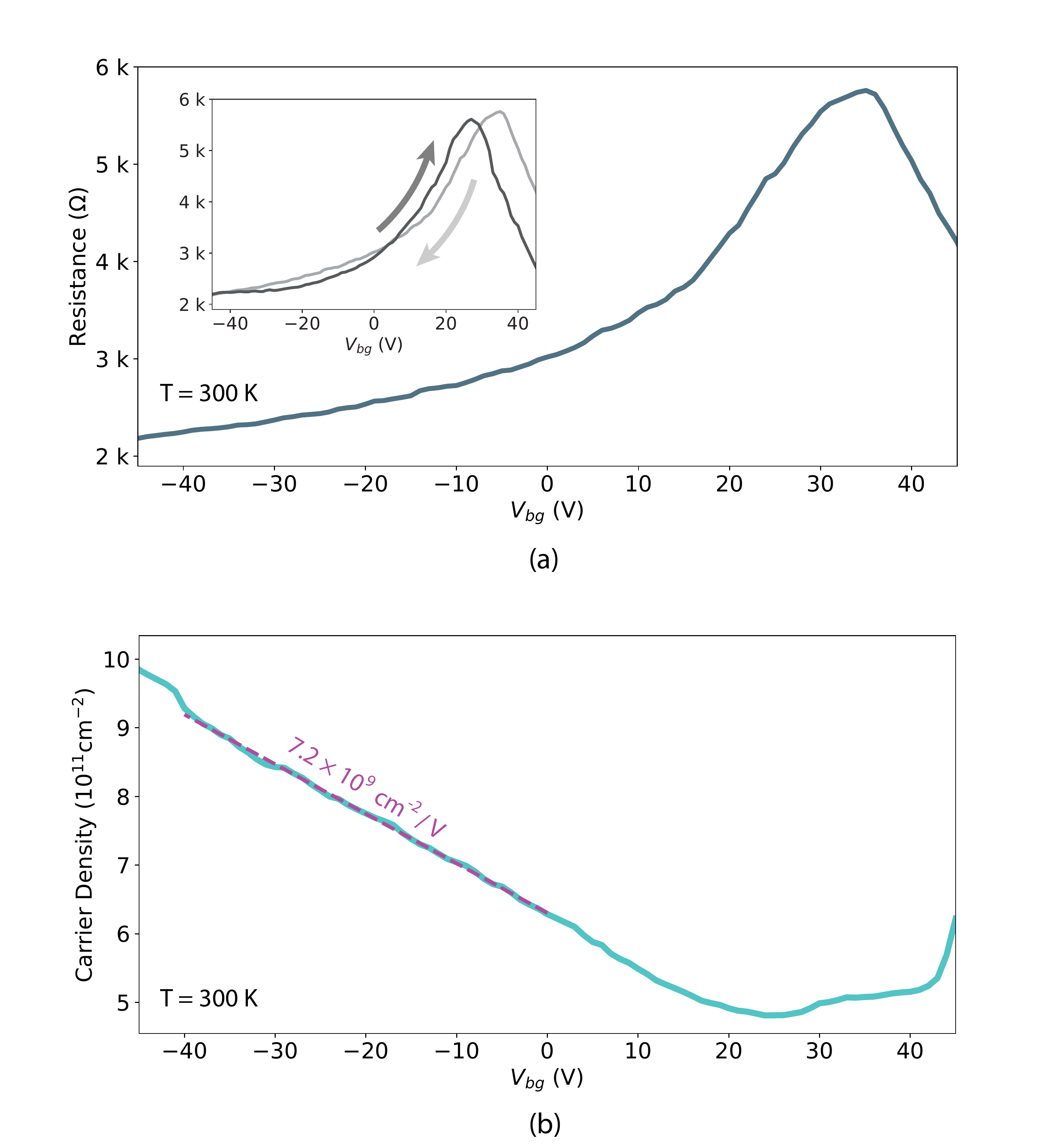}
\label{FIGS1}
\caption{(a) shows the resistance of graphene as a function of backgate voltage at $T = 300$ K. The measurement is performed before the c-AFM writing as a control measurement. The four-terminal resistance is measured in the same configuration as in Figure 1(c) of main text. The inset of (a) shows the hysteretic behavior of graphene on LAO/STO. The CNP for backward backgate sweeping is at $V_\mathrm{bg} = +35$ V, while the for forward backgate sweeping, CNP is at  $V_\mathrm{bg} = +28$ V. (b) Carrier density measured from Hall effect, with $V_\mathrm{bg}$ tuned in the same range as in (a). The carrier density $n$ vs. $V_\mathrm{bg}$ data is fit to a linear model for $V_\mathrm{bg}$ between $-40$ V and $0$ V, with a slope $k = 7.2 \times 10^9 $ cm$^{-2}$/V. For $V_\mathrm{bg}$ between $+20$ V and $+45$ V, the device is close to the CNP and the carrier density is not well-defined due to the existence of electron-hole puddles.}
\end{figure*}

\begin{figure*}[tp]
\includegraphics[width=1\textwidth]{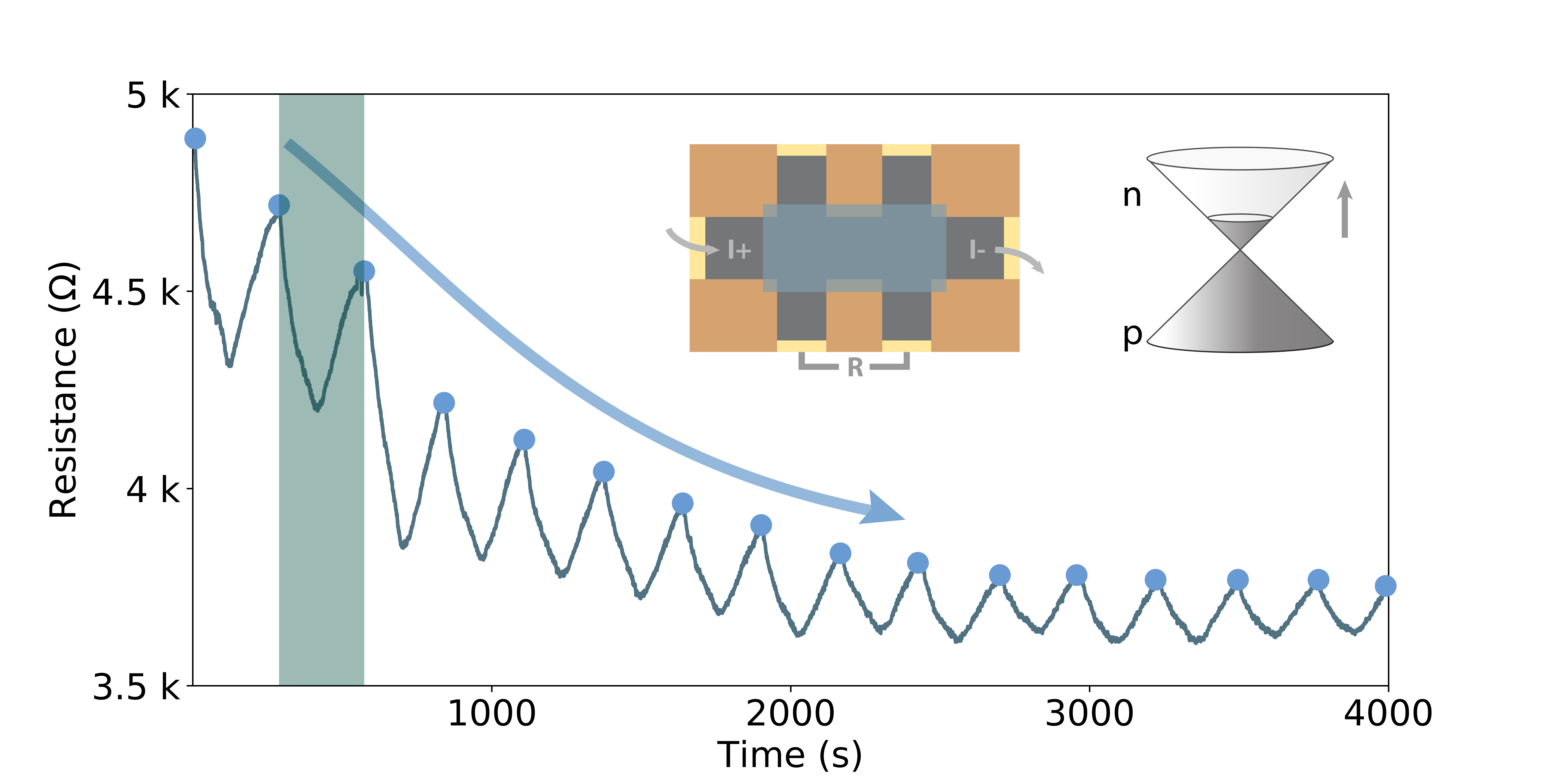}
\label{FIGS2}
\caption{Four-terminal resistance of graphene during c-AFM lithography performed at room temperature.  The resistance of the graphene changes when continuous raster scans are performed, indicating an overall n-type doping of the graphene over time. The scanning region is illustrated in the inset.  For this experiment, the entire graphene region is written by c-AFM. Blue dots mark the beginning/ending of consecutive AFM raster scans. 
The AFM scanning speed is 10 $\mu$m/s, and it takes 260 s to scan the graphene device (5 $\mu$m $\times$ 10 $\mu$m). 
The green region marks one complete scan. The recovery of resistance and doping level is similar to the conductivity decay mechanism of nanowires written by c-AFM on bare LAO/STO. }
\end{figure*}

\begin{figure*}[tp]
\includegraphics[width=0.6\textwidth]{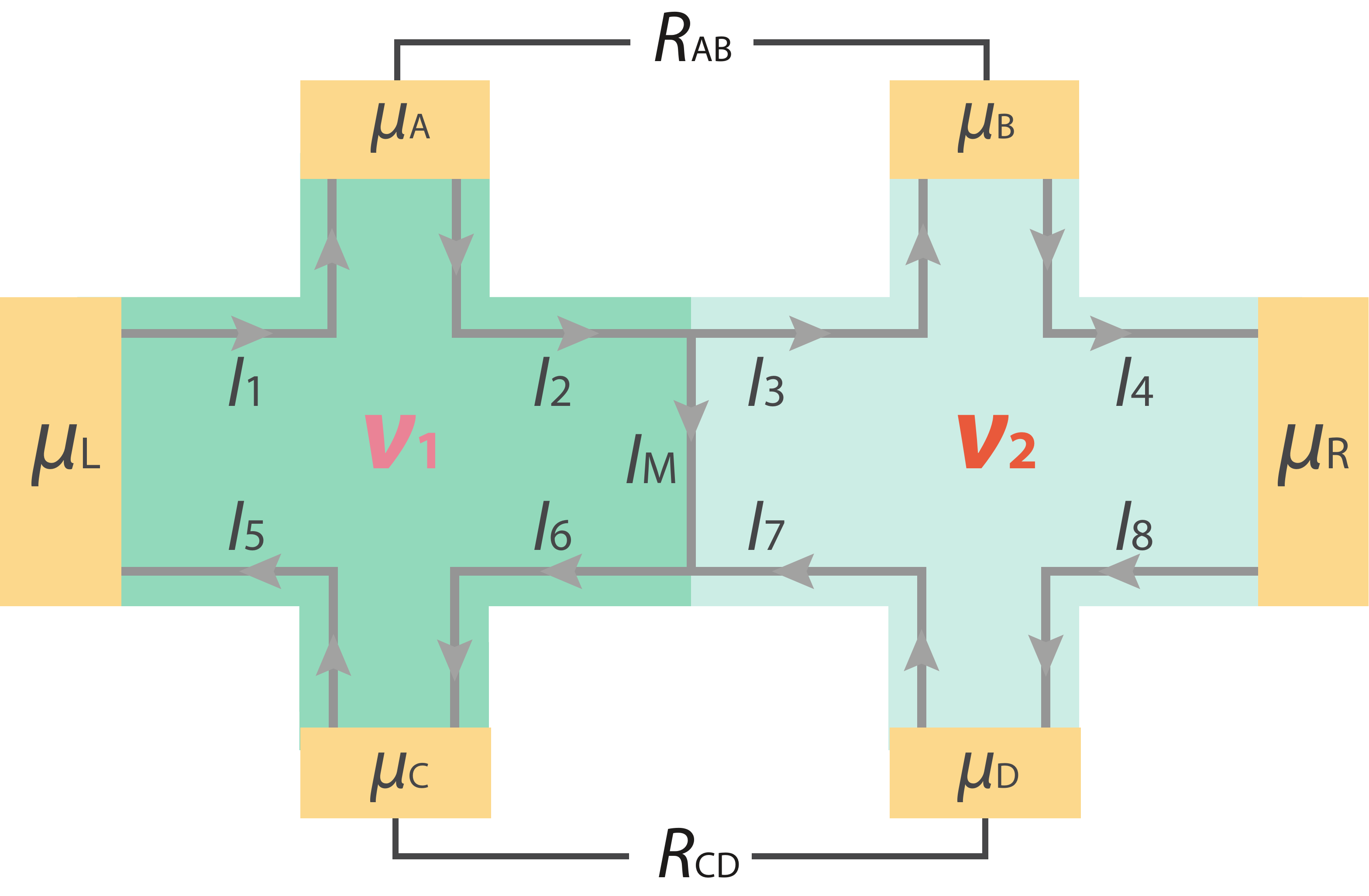}
\label{FIGS3}
\caption{Derivation for the longitudinal resistance $R_{AB}$ and $R_{CD}$ for unipolar case. Assume the current is sourced from the left hand side. $\mu_L$ and $\mu_R$ are the chemical potentials of the source and drain electrodes. $\mu_A$, $\mu_B$, $\mu_C$ and $\mu_D$ are the chemical potentials of the voltage leads. When the two regions are unipolar, with $|\nu_1| > |\nu_2|$, the edge currents are $\displaystyle I_1 = \frac{e}{h}\mu_L|\nu_1|$, $\displaystyle I_2 = \frac{e}{h}\mu_A|\nu_1|$, $\displaystyle I_3 = I_2 - I_M$, $\displaystyle I_4 = \frac{e}{h}\mu_B|\nu_2|$, $\displaystyle I_5 = \frac{e}{h}\mu_C|\nu_1|$, $\displaystyle I_6 = I_M + I_7$, $\displaystyle I_7 = \frac{e}{h}\mu_D|\nu_2|$, $\displaystyle I_8 = \frac{e}{h}\mu_R|\nu_2|$. The left hand side has a higher Landau level filling factor and more edge channels, so the channels with higher energy levels in $I_2$ are reflected into $I_M$, therefore $\displaystyle I_M = \frac{e}{h}\mu_L(|\nu_1| - |\nu_2|)$. The net currents flowing into and out of voltages leads A, B, C and D are zero, so $I_1 = I_2$, $I_3 = I_4$, $I_5 = I_6$ and $I_7 = I_8$, and therefore $\mu_A = \mu_L$, $\mu_B = \mu_L$, $\displaystyle \mu_C = \mu_L + \frac{|\nu_2|}{|\nu_1|}(\mu_L - \mu_R)$, $\mu_D = \mu_R$. The total current from the source is $\displaystyle I = I_1 - I_5 = \frac{e}{h}(\mu_L - \mu_R)|\nu_1|$. Finally, the longitudinal resistance $\displaystyle R_{AB} = \frac{\mu_A - \mu_B}{eI} = 0$, and $\displaystyle R_{CD} = \frac{\mu_C - \mu_D}{eI} = \frac{h}{e^2}\left(\frac{1}{|\nu_2|} - \frac{1}{|\nu_1|}\right)$.}
\end{figure*}

\begin{figure*}[tp]
\includegraphics[width=0.6\textwidth]{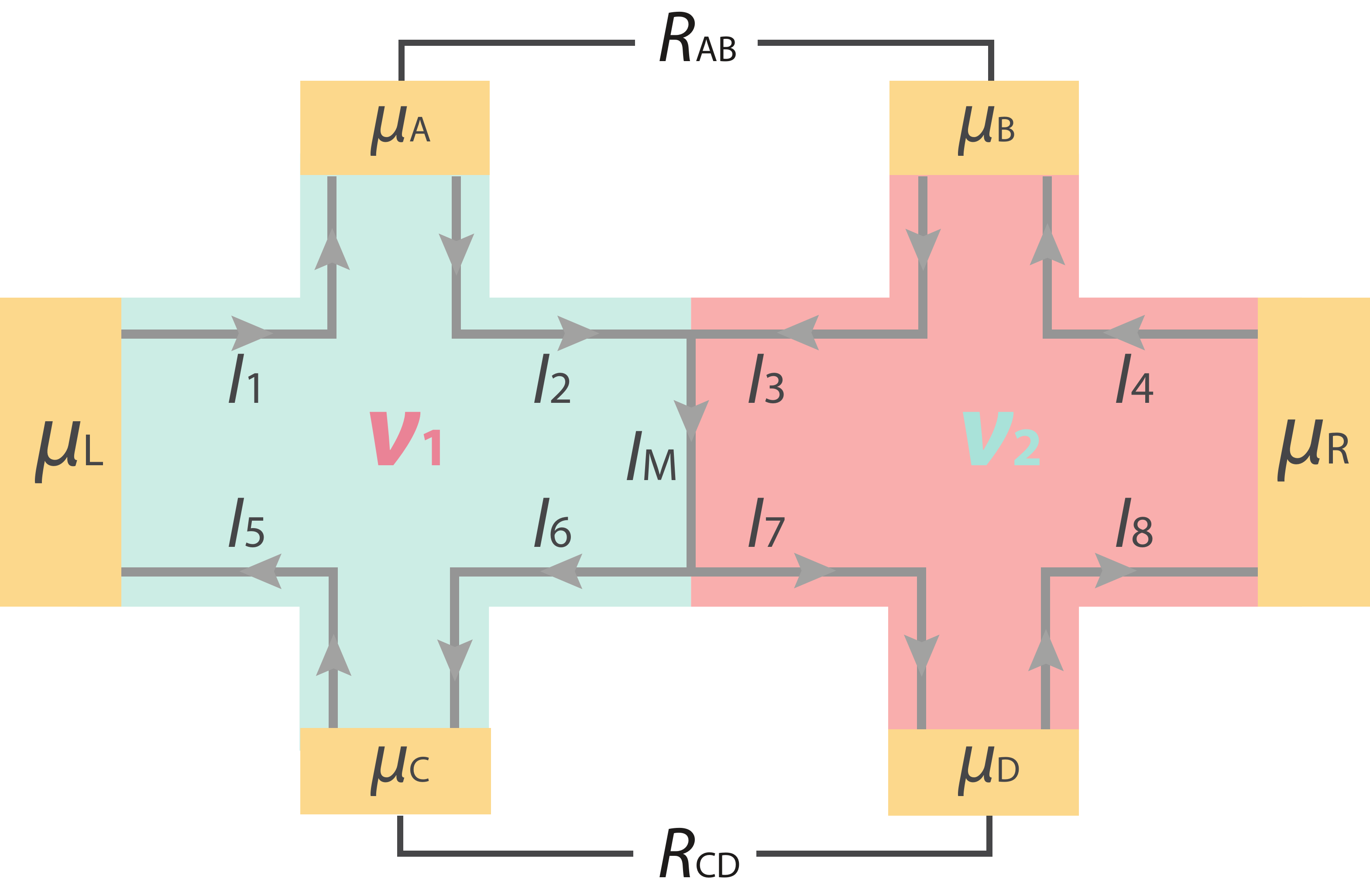}
\label{FIGS4}
\caption{Derivation for the longitudinal resistance $R_{AB}$ and $R_{CD}$ for bipolar case. Assume the current is sourced from the left hand side. $\mu_L$ and $\mu_R$ are the chemical potentials of the source and drain electrodes. $\mu_A$, $\mu_B$, $\mu_C$ and $\mu_D$ are the chemical potentials of the voltage leads. When the two regions are bipolar, the currents in the two regions are flowing in opposite directions. The edge currents are $\displaystyle I_1 = \frac{e}{h}\mu_L|\nu_1|$, $\displaystyle I_2 = \frac{e}{h}\mu_A|\nu_1|$, $\displaystyle I_3 = \frac{e}{h}\mu_B|\nu_2|$, $\displaystyle I_4 = \frac{e}{h}\mu_R|\nu_2|$, $\displaystyle I_5 = \frac{e}{h}\mu_C|\nu_1|$, $\displaystyle I_8 = \frac{e}{h}\mu_D|\nu_2|$. Assume the edge channels are fully mixed on the boundary of the two regions, $\displaystyle I_M = I_2 + I_3$, $\displaystyle I_6 = \frac{|\nu_1|}{|\nu_1|+|\nu_2|} \cdot I_M$ and $\displaystyle I_7 = \frac{|\nu_2|}{|\nu_1|+|\nu_2|} \cdot I_M$. The net currents flowing into and out of voltages leads A, B, C and D are zero, so $I_1 = I_2$, $I_3 = I_4$, $I_5 = I_6$ and $I_7 = I_8$, and therefore $\mu_A = \mu_L$, $\mu_B = \mu_R$, $\displaystyle \mu_C = \mu_D = \frac{\mu_L|\nu_1| + \mu_R|\nu_2|}{|\nu_1| + |\nu_2|}$. The total current from the source is $\displaystyle I = I_1 - I_5 = \frac{e}{h} \cdot \frac{|\nu_1||\nu_2|}{|\nu_1|+|\nu_2|} \cdot (\mu_L - \mu_R)$. Finally, the longitudinal resistance $\displaystyle R_{AB} = \frac{\mu_A - \mu_B}{eI} = \frac{h}{e^2}\left(\frac{1}{|\nu_1|} + \frac{1}{|\nu_2|}\right)$, and $\displaystyle R_{CD} = \frac{\mu_C - \mu_D}{eI} = 0$.}
\end{figure*}